\documentclass[preprint2]{aastex}

\slugcomment{To appear in Dec. 20 issue of Astrophys. J.}

\shorttitle{Dust Radial Flow}
\shortauthors{Takeuchi \& Lin}

\begin{document}


\title{Radial Flow of Dust Particles in Accretion Disks \footnote{To
appear in Dec. 20 issue of Astrophys. J.}}

\author{Taku Takeuchi and D. N. C. Lin}
\affil{UCO/Lick Observatory, University of California, Santa Cruz,
CA95064}
\email{taku@ucolick.org, lin@ucolick.org}

\begin{abstract}
We study the radial migration of dust particles in accreting
protostellar disks analogous to the primordial solar nebula.  Our
main objective is to determine the retention efficiency of dust
particles which are the building blocks of the much larger
planetesimals. This study takes account of the two dimensional
(radial and normal) structure of the disk gas, including the effects of
the variation in the gas velocity as a function of distance from the
midplane.
It is shown that the dust component of disks accretes slower than the
gas component. At high
altitude from the disk midplane (higher than a few disk scale heights),
the gas rotates faster than particles because of the inward pressure
gradient force, and its drag force causes particles to move outward in the
radial direction.  Viscous torque induces the gas within a
scale height from the disk midplane to flow outward, carrying small
(size $\la 100 \ \micron$ at $10 \ {\rm AU}$) particles with it.
Only particles at intermediate altitude or with sufficiently large
sizes ($\ga 1 \ {\rm mm}$ at $10 \ {\rm AU}$) move inward.  When the
particles' radial velocities are averaged over the entire
vertical direction, particles have a net inward flux.  The magnitude
of their radial motion depends on the particles' distance from the
central star.  At large distances, particles migrate inward with a
velocity much faster than the gas accretion velocity.  However, their
inward velocity is reduced below that of the gas in the inner regions
of the disk.  The rate of velocity decrease is a function of the
particles' size.  While larger particles retain fast accretion
velocity until they approach closer to the star, $10 \ \micron$
particles have slower velocity than the gas in the most part
of the disk ($r \la 100$ AU).  This differential migration of
particles causes the size fractionation.  Dust disks composed mostly
of small particles (size $\la 10 \ \micron$) accrete slower than gas
disks, resulting in the increase in the dust-gas ratio during the gas
accretion phase.  If the gas disk has a steep radial density gradient
or if dust particles sediment effectively to the disk midplane,
the net vertically averaged flux of particles can be
outward. In this case, the accretion of the dust component is
prevented, leading to the formation of residual dust disks after
their gas component is severely depleted.

\end{abstract}



\section{Introduction}

Planets form in circumstellar disks.  In the standard scenario,
formation of earth-size planets or planetary cores occurs through
coagulation of small dust particles (e.g., Weidenschilling \& Cuzzi
1993).  Thus, the total amount, radial density distribution, and size
distribution of dust particles in the disks are the important initial
conditions of planet formation.  In many of the previous studies of
planet formation, the dust-gas ratio of circumstellar disks is assumed to
be similar to the solar value ($\sim 10^{-2}$), and dust particles are
considered to be well mixed with gas, i.e., the dust-gas ratio is
constant throughout the disk (Hayashi, Nakazawa, \& Nakagawa 1985).

However, the motion of dust particles is different from that of gas.
Dust particles have a radial motion which is induced by the gas drag
force.  Adachi, Hayashi, \& Nakazawa (1976) and Weidenschilling (1977)
studied the motion of particles in gas disks.  Due to radially outward
pressure gradient force, the rotation velocity of the gas is generally
slower than particles on nearly Keplerian circular orbits. Consequently, 
the gas drag force takes angular momentum from particles, resulting in
their inward migration (Whipple 1972).  At a few AU from the central
star, the orbital decay time is estimated to be $\sim 10^6 \ {\rm yr}$
for $100 \ \micron$ particles and $10^4 \ {\rm yr}$ for $1 \ {\rm cm}$
particles. 
These times are much shorter than the life time of gas
disks ($\sim 10^7 \ {\rm yr}$).  Thus, dust component may evolve much
faster than the gas in protostellar disks, in which case, the dust-gas
ratio would decline.  It appears that the initial conditions of
planet formation need not to be protostellar disks in which dust
particles are well mixed with the gas and the dust-gas ratio is constant.

In this paper, we start a series of studies to determine the initial
distribution of dust particles in the context of planet formation.
Here, we study the radial migration of dust particles in gas disks.  We
focus our attention on particles smaller than $\sim 1 \ {\rm cm}$ in
order to consider evolution of dust disks before the initiation of
planetesimal formation.  The migration of these small particles
establishes the initial density distribution of the dust component and
sets the stage for planetesimal formation. In this phase of disk
evolution, the gas component is considered to be turbulent and is
accreting onto the star.  The discussion in the above paragraph
is based on the study of particle migration under the assumption that
particles have completely sedimented on the midplane of laminar disks.
However, in turbulent disks, the sedimentation of particles is
prevented, because particles are stirred up to high altitude from the
midplane by turbulent gas motion.  In such disks, particles are
distributed in the vertical direction, and the radial motion of
particles depends on the distance from the midplane.  Thus, there is a
need for studying the radial migration of particles which reside above
the midplane.

There are two important factors causing the vertical variation in the
migration velocity of particles.  The first is the variation of
rotation velocity of the gas in the vertical direction.  As mentioned
above, the gas rotation differs from the Keplerian rotation because of
the gas pressure 
gradient.  As the gas density decreases with the distance from the
midplane, the radial pressure gradient varies and under some circumstances
even changes its sign.  While at the midplane the pressure gradient
force is
outward, it is inward near the disk surface because the disk thickness
increases with the radius.  The gas drag force on particles also varies
with the height, resulting in a variation of particle migration
velocities.  In \S \ref{sec:vrd}, we will see that particles at high
altitude, where the pressure gradient force is inward, flow outward.

The second factor is the variation in the radial gas flow.
Small particles ($\la 100 \ \micron$) are well coupled to the gas, so
that they migrate with the gas flow, as discussed in \S \ref{sec:vrd}
below.  That is if the gas is accreted onto the star, the small particles
would also be accreted, while if the gas flows outward, particle flow
would also be outward.  In accreting protostellar disks, the gas does not
always flow radially inward.  Radial velocity of the gas varies with
the distance from the midplane.  Viscous stress can cause gas outflow
near the midplane (Urpin 1984; Kley \& Lin 1992).
R\'o\.zyczka, Bodenheimer, \& Bell (1994) showed that
the outflow occurs if the radial pressure gradient at the midplane is
steep enough.  The density (or pressure) gradient at the midplane is
steeper than the average spatial (or surface) density gradient, 
e.g.,  if the surface density varies as $\Sigma_g \propto r^{-1}$ and
the disk thickness varies as $h_g \propto r$, the midplane density 
has a steeper variation corresponding to $\rho_g \propto r^{-2}$.
Viscous diffusion at the 
midplane causes the outflow of the gas to reduce the steep density gradient.
On the other hand, at high altitude from the midplane, the density
gradient is shallow and the viscous torque causes the usual inward flow.
Thus, in accretion disks, the outward flow at the midplane is
sandwiched by the inflow at the disk surfaces (see \S\ref{sec:vrg} for
details).
Because small particles are carried by the gas flow, they are expected
to flow outward near the midplane, and inward at higher altitude.

Thus, if most of dust particles concentrate in the region where they
migrate outward, i.e., near the
disk surface or around the midplane, the dust component of the disk
would accumulate in total mass and expand in size. In reality, the
vertical spatial distribution of particles is determined by the
degree of sedimentation, and their size distribution is regulated by
the coagulation, condensation, and sublimation processes. In this
paper, we calculate the vertical distribution of particles in order to
identify the mass flux of the dust particles as functions of their
size and their distance from the midplane and their host stars.  The
results of the present calculation will be used to study the long term
dynamical evolution of the dust component of the disk. For the present
task, we do not consider the effects of gas depletion, particles' size
evolution or their feedback influence on the flow velocity of the gas.
These effects will be considered in a future investigation.

There are several works on the particle migration in turbulent disks.
Stepinski \& Valageas (1997) derived the vertically averaged velocity
of migration for two limiting cases. One model is constructed for
small particles which are well mixed with the gas and the other is for
large particles which are totally sedimented to the midplane. The
velocity for intermediate sized particles is calculated by the
interpolation of these limiting cases.  Supulver \& Lin (2000)
investigated the evolution of particle orbits in turbulent disks
numerically.  This paper gives analytical expression of the particle
velocity for any size of particles, which are estimated from
interpolation by Stepinski \& Valageas (1997).  Several studies showed
that vortices or turbulent eddies can trap particles within some size
range (Barge \& Sommeria 1995; Cuzzi, Dobrovolskis, \& Hogan 1996; Tanga
et al. 1996; Klahr \& Henning 1997; Cuzzi et al. 2001, see however
Hodgson \& Brandenburg 1998).  Particle trapping
by eddies may be important for local 
collection of size-sorted particles.  In this paper, turbulent
eddies are treated just as a source of the particle diffusion and of the
gas viscosity.  The effect of particle trapping on their coagulation is 
not discussed here.

The plan of the paper is as follows.  In \S 2, the vertical variation
of gas flow is derived.  In \S 3, we describe how the radial velocity
of dust particles varies in the vertical direction, and then calculate
the net radial migration velocity.  In \S 4, we discuss the steady
distribution of particles assuming no particle growth. We show that
the size fractionation of particles occurs as a result of radial
migration, and that the distribution of particles differs from
that of the gas.


\section{Viscous Flows of Gas Disks}

\subsection{Rotation Law of the Gas}

Gas disks rotate with a slightly different velocity from the Keplerian
velocity.  The rotation velocity is determined by the balance between
the stellar gravity, the centrifugal force, and the gas pressure gradient.
In cylindrical coordinates $(r,z)$, the balance of the forces
described by the momentum equation in the $r$-direction is given by
\begin{equation}
r \Omega_g^2 - \frac{G M r}{(r^2+z^2)^{3/2}} - \frac{1}{\rho_g}
\frac{\partial P_g}{\partial r} = 0 \ ,
\label{eq:gaseq_r}
\end{equation}
where $\Omega_g$ is the angular rotation velocity of the gas, $G$ is the
gravitational constant, $M$ is the stellar mass, $\rho_g$ is the gas
density, and $P_g$ is the gas pressure.  The density distribution in
the vertical direction is determined by the balance between the
$z$-components of the stellar gravity and the pressure gradient, which
is written as
\begin{equation}
- \frac{G M z}{(r^2+z^2)^{3/2}} - \frac{1}{\rho_g} \frac{\partial
P_g}{\partial z} = 0 \ .
\end{equation}
We assume that the disk is isothermal in the vertical direction.
Neglecting $(z/r)^2$ and higher order terms, the vertical density
distribution is obtained as
\begin{equation}
\rho_g (r,z) = \rho_g (r,0) \exp \left( - \frac{z^2}{2 h_g^2} \right) \ .
\label{eq:gasdensity}
\end{equation}
The disk scale height $h_g$ is given by
\begin{equation}
h_g (r) = \frac{c}{\Omega_{\rm K,mid}} \ ,
\label{eq:hg}
\end{equation}
where $c$ is the isothermal sound speed, and $\Omega_{\rm K,mid} =
\sqrt{GM/r^3}$ is the Keplerian angular velocity at the midplane.  The
radial variations of physical quantities are assumed to have power law
forms, such that
\begin{eqnarray}
\rho_g (r,z) & = & \rho_0 r_{\rm AU}^p \exp \left( - \frac{z^2}{2 h_g^2}
\right) \ , \nonumber \\
c^2 (r) & = & c_0^2  r_{\rm AU}^q \ , \nonumber \\
h_g (r) & = & h_0  r_{\rm AU}^{(q+3)/2} \ ,
\label{eq:powerlaw1}
\end{eqnarray}
where the subscript ``0'' means the values at $1 \ {\rm AU}$, $r_{\rm AU}$
is the radius in the unit of AU, and power law indices $p$ and $q$ are
usually negative.  The surface density is
\begin{equation}
\Sigma_g (r) = \int_{-\infty}^{+\infty} \rho_g dz
= \sqrt{2\pi} \rho_0 h_0 r_{\rm AU}^{p_s} \ ,
\end{equation}
where $p_s=p+(q+3)/2$.
To derive the rotation law of the gas, we retain $(z/r)^2$ and lower order
terms in equation (\ref{eq:gaseq_r}), then obtain
\begin{equation}
\Omega_g (r,z) = \Omega_{\rm K,mid} \left[ 1 + \frac{1}{2} \left(
\frac{h_g}{r} \right)^2 \left( p+q+ \frac{q}{2} \frac{z^2}{h_g^2}
\right) \right] \ .
\label{eq:omegag}
\end{equation}
The difference of the gas rotation from the Kepler rotation is order
of $(h_g/r)^2$, and the gas at high altitude rotates slower than the gas at
the midplane.

\subsection{Radial Velocity of the Gas Flow \label{sec:vrg}}

In viscous disks, angular momentum is transfered from the inner part to
the outer part of the disk under the action of viscous stress.  The azimuthal
component of the momentum equation implies 
\begin{eqnarray}
2 \pi r \rho_g \left( v_{r,g} \frac{\partial}{\partial r} + v_{z,g}
\frac{\partial}{\partial z} \right) (r^2 \Omega_g) = \nonumber \\
2 \pi \left[
\frac{\partial}{\partial r} \left( r^3 \rho_g \nu \frac{\partial
\Omega_g}{\partial r} \right) + \frac{\partial}{\partial z} \left( r^3
\rho_g \nu \frac{\partial \Omega_g}{\partial z} \right)\right] \ ,
\label{eq:gasang}
\end{eqnarray}
where $v_{r,g}$ and $v_{z,g}$ are the $r$- and $z$-components of the gas
velocity, respectively, and $\nu$ is the kinetic viscosity.  The right
hand side of equation (\ref{eq:gasang}) represents the torque exerted
on an annulus with unit width in the $r$- and $z$-directions.  The
left hand side is the variation of the angular momentum of the annulus
accompanying with the motion.  We assume that the angular velocity is
always adjusted to the equilibrium (eq. [\ref{eq:omegag}]) on a
dynamical time scale.  Since molecular viscosity is too ineffective,
it is customary to assume that angular momentum is primarily
transported by the Reynolds stress induced by the turbulent motion of
the gas.  Following conventional practice, we adopt an {\it ad hoc} but
simple to use $\alpha$ prescription for turbulent viscosity (Shakura
\& Sunyaev 1973), such that
\begin{equation}
\nu = \alpha c h_g = \alpha c_0 h_0 r_{\rm AU}^{q+3/2} \ .
\label{eq:nu}
\end{equation}
The mass conservation of the gas is
\begin{equation}
\frac{\partial \rho_g}{\partial t} + \frac{1}{r}
\frac{\partial}{\partial r} (r \rho_g v_{r,g}) + \frac{\partial}{\partial
z} (\rho_g v_{z,g} ) = 0 \ .
\label{eq:gascont1}
\end{equation}
In accretion disks considered here, the second term is the same order
of the third term.  Then, we see that $v_{z,g} \sim (h_g/r) v_{r,g}$.
Besides, from equation (\ref{eq:omegag}) we see $\partial (r^2 \Omega_g)
/ \partial z \sim (h_g/r) \partial
(r^2 \Omega_g) / \partial r$.  Thus, $v_{z,g} \partial (r^2 \Omega_g)
/ \partial z$ is factor $(h_g/r)^2$ smaller than $v_{r,g} \partial
(r^2 \Omega_g) / \partial r$ and can be neglected in equation
(\ref{eq:gasang}).  Using power law expressions (\ref{eq:powerlaw1})
and (\ref{eq:nu}) and retaining terms of order $(h_g/r)^2$, the radial
velocity is reduced to
\begin{eqnarray}
v_{r,g} & = & -2 \pi \alpha \left( \frac{h_0}{{\rm AU}} \right)^2 \left( 3p
+ 2q + 6 + \frac{5q+9}{2} \frac{z^2}{h_g^2}
\right) \nonumber \\
& & \times \ r_{\rm AU}^{q+1/2} \ {\rm AU \ yr}^{-1} \ .
\label{eq:vrg}
\end{eqnarray}

In the following discussions, we adopt these values as being
representative of a typical model: $M=1 M_{\sun}$, $\rho_0 = 2.83 \times
10^{-10} \ {\rm g \ cm}^{-3}$, $h_0 = 3.33 \times 10^{-2} \ {\rm AU}$,
$p=-2.25$, 
$q=-0.5$, and $\alpha = 10^{-3}$.  These values correspond to a
minimum mass solar nebula whose surface density distribution has a
power law form with index $p_s = p+(q+3)/2 = -1.0$ and has a gas mass
$2.5 \times 10^{-2} M_{\sun}$ inside 100AU.

\begin{figure}
\epsscale{1.0}
\plotone{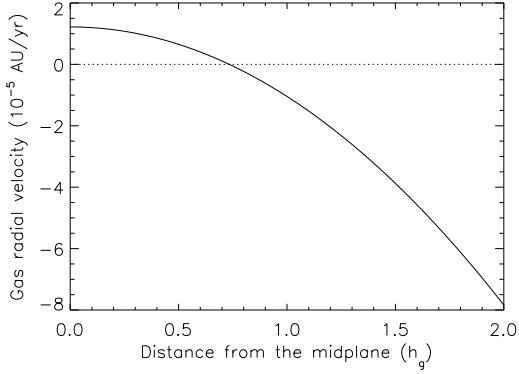}
\caption{
Radial velocity of gas $v_{r,g}$. The distance from the midplane $z$ is
normalized by the disk scale height $h_g$. In the standard model
($q=-0.5$), the radial velocity is independent of the distance from the
star. The dotted line shows the zero velocity for reference.
\label{fig:vrg}
}
\end{figure}

Figure \ref{fig:vrg} shows the radial velocity of the gas.  The radial
velocity is outward (positive) near the midplane, while it is inward
(negative) above the height $0.73 h_g$.  In the standard model
($q=-0.5$), the radial velocity is independent of the distance from the
star, as seen from equation (\ref{eq:vrg}).  At the midplane, the radial
gradient of the gas density is so steep ($p=-2.25$) that the gas receives
more torque from 
the inner disk than the torque it exerts on the outer disk.  This
radial dependence means that viscous stress acts to move the gas outward
in such a steep density gradient.  The condition for the outward gas
flow at the midplane is
\begin{equation}
p + \frac{2}{3} q < -2 \ ,
\label{eq:condoutflow}
\end{equation}
which is satisfied in many accretion disk models.  The rotation of the gas
is slower at higher altitude as seen in equation (\ref{eq:omegag}).
Thus, the gas at the midplane loses its angular momentum through the
action of viscous stress it has with the gas at higher altitude.  However,
in the standard model, this angular momentum loss is smaller than the angular
momentum gain through the viscous stress in the $r$-direction.  The
outflow zone around the midplane shrinks as the radial density
gradient is reduced ($p \rightarrow 0$), and when the condition
(\ref{eq:condoutflow}) is 
violated, gas at all altitudes flows inward.  The radial density
gradient is shallower at higher altitude, because the disk
scale height increases with the radius.  Thus, at high altitude, the gas
loses more angular momentum through exerting torque on the outer disk
than it is receiving from the inner disk.  For such
gas, the presence of viscous torque induces it to flow towards the star.

The net radial velocity of the gas is given by averaging in the vertical
direction:
\begin{eqnarray}
\langle v_{r,g} \rangle & = & \frac{1}{\Sigma_g}
\int_{-\infty}^{+\infty} v_{r,g} \rho_g dz \nonumber \\
& = &  -6 \pi \alpha \left( \frac{h_0}{\rm AU} \right)^2 (p_s +q+2)
\nonumber \\
& & \times \ r_{\rm AU}^{q+1/2} \ {\rm AU \ yr^{-1}} \ .
\end{eqnarray}
The net accretion of the gas toward the star occurs if
\begin{equation}
p_s + q = p + \frac{3}{2} (q+1) > -2 \ .
\label{eq:condaccretion}
\end{equation}
We may consider two models in which a steady state is achieved.  One
model is the case in which the net radial velocity is zero, i.e., $p_s + q =
-2$.  For example, if $p_s=-1.5$ and $q=-0.5$, there is no net accretion
of the disk. A more realistic model is the case in which the mass
flux of the gas is constant with the radius, i.e., $r \Sigma_g \langle
v_{r,g} \rangle \propto r^{p_s + q + 3/2}$ is constant.  We take this
as our standard model, i.e., $p_s=-1.0$ and $q=-0.5$.  This surface
density profile is expected to arise after the disk has undergone a
period of initial viscous evolution.


\section{Radial Flow of Dust Particles}

\subsection{Vertical Dependence of Radial Velocity \label{sec:vrd}}

The azimuthal velocities of dust particles are different from that of
the gas.  The resulting gas drag force transfers angular momentum
between the particles 
and the gas and moves particles in the radial direction.  If there is no gas
drag force, particles would orbit with the Keplerian angular velocity, which
is approximated as
\begin{equation}
\Omega_{\rm K} (r,z) \approx \Omega_{\rm K,mid} \left( 1 - \frac{3}{4}
\frac{z^2}{r^2} \right) \ .
\label{eq:omegak}
\end{equation}
We express the deviation of the angular velocity of the gas from the
Keplerian angular velocity as
\begin{equation}
\Omega_g = \Omega_{\rm K} (1-\eta)^{1/2} \ .
\label{eq:veldiff}
\end{equation}
From equation (\ref{eq:gaseq_r}) it is seen that \\
$\eta=-(r \Omega_{\rm
K}^2 \rho_g)^{-1} \partial P_g / \partial r$ is the ratio of the gas
pressure gradient to the stellar gravity in the radial direction.
From equations (\ref{eq:omegag}), (\ref{eq:omegak}), and
(\ref{eq:veldiff}), $\eta$ is written within the order of $(h_g/r)^2$ as,
\begin{equation}
\eta = -\left( \frac{h_g}{r} \right)^2 \left( p+ q+ \frac{q+3}{2}
\frac{z^2}{h_g^2} \right) \ .
\label{eq:eta}
\end{equation}
Note that the pressure gradient force is outward ($\eta$ is positive) around
the midplane, while it is inward ($\eta$ is negative) where $|z| >
\sqrt{-2(p+q)/(q+3)} h_g$.  In the standard model, $\eta$ changes
sign at $z \approx 1.5 h_g$.  Dust particles near the midplane rotate
faster than the gas, and particles at high altitude rotate slower than
the gas.

The equations of motion of a particle are
\begin{equation}
\frac{d v_{r,d}}{dt} = \frac{v_{\theta,d}^2}{r} - \Omega_{\rm K}^2 r -
\frac{\Omega_{\rm K,mid}}{T_s} ( v_{r,d} - v_{r,g} ) \ ,
\label{eq:motion_r}
\end{equation}
\begin{equation}
\frac{d}{dt} \left( r v_{\theta,d} \right) =
 - \frac{v_{\rm K,mid}}{T_s} ( v_{\theta,d} - v_{\theta, \rm g} ) \ ,
\label{eq:motion_th}
\end{equation}
where $v_r$ and $v_{\theta}$ are the $r$- and $\theta$- components of
the velocity, respectively, with the subscripts ``$g$'' and ``$d$''
distinguishing gas and dust, and $v_{\rm K,mid}=r \Omega_{\rm K,mid}$ is the
Keplerian velocity at the midplane.  The gas drag force is expressed
through the non-dimensional stopping time $T_s$ (normalized by the
Kepler time at the midplane).
We do not solve the equation of motion in the $z$-direction. Instead, we
simply assume $v_{z,d}=0$. When an equilibrium in the vertical dust
distribution is achieved, the dust sedimentation to the midplane is
balanced by the diffusion of particles, as discussed below in
\S\ref{sec:vdist}. The vertical velocity of a particle is zero when time
averaged.

The mean free path
of gas molecules is larger than $1 \ {\rm cm}$ for $r \ga 1 \ {\rm AU}$
in our models (Nakagawa, Sekiya, \& Hayashi 1986).  In
this paper, we consider particles smaller than $1 \ {\rm cm}$ and use
Epstein's gas drag law.  Then, the non-dimensional stopping time $T_s$,
as given by Takeuchi \& Artymowicz (2001), is
\begin{equation}
T_{s} = \frac{\rho_{p} s v_{\rm K,mid}}{\rho_{g} r v_{T}} \ ,
\label{eq:stoptime}
\end{equation}
where $\rho_p$ is the particle internal density, $s$ is the particle
radius, and the mean thermal velocity is $v_T =\sqrt{8 / \pi} c$
\footnote{In Takeuchi \& Artymowicz (2001), $v_T$ is defined to be
$4/3$ times the mean thermal velocity. This definition causes a factor
$4/3$ difference between equation (\ref{eq:stoptime}) and their
equation (10).}.  We take the particle internal density as
$\rho_p=1.25 \ {\rm g \ cm^{-3}}$.

\begin{figure}
\epsscale{1.0}
\plotone{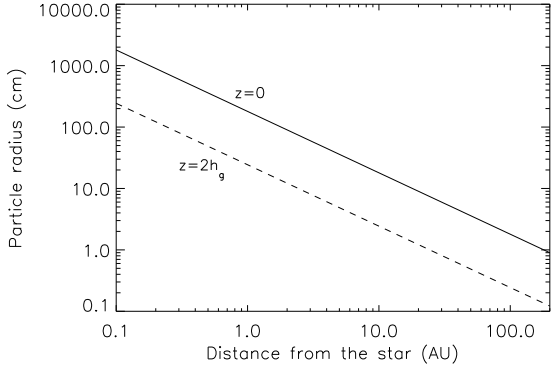}
\caption{
The particle radius whose non-dimensional stopping time $T_s$ is unity
in the standard disk ($p_s=-1.0$ and $q=-0.5$).  The solid line
corresponds to particles at the midplane, while the dashed line
corresponds to particles at altitude $z=2h_g$. Particles smaller than the
lines have $T_s$ smaller than unity and couple well with the gas, while
particles larger than the lines have $T_s >1$.
\label{fig:stoptime}
}
\end{figure}

For particles smaller than $1 \ {\rm cm}$, the non-dimensional
stopping time is much smaller than unity through most of the disk.
Figure \ref{fig:stoptime} shows the radii of particles whose stopping
time is unity. At the midplane (see the solid line), the non-dimensional
stopping time is 
smaller than unity for particles of size $s \la 1 \ {\rm cm}$ at $r
\la 100 \ {\rm AU}$.  At $z=2h_g$ (the dashed line), although the gas
density is lower
than at the midplane and the stopping time is longer, particles of
size $s \la 1 \ {\rm mm}$ still have the non-dimensional stopping time
smaller than unity.  These particles are well coupled to the gas and
their angular velocity is similar to the gas angular velocity
(\ref{eq:omegag}).  Only particles with $s \ga 1 \ {\rm cm}$ which are
located at high altitude and $r \ga 100 \ {\rm AU}$ become decoupled
from the gas.

We assume that motions of both the gas and the particles are close to
Keplerian, i.e., $v_{\theta,g} \approx v_{\theta,d} \approx v_{\rm
K,mid}$, and that $d(r v_{\theta,d})/dt \approx v_{r,d} d(r v_{\rm
K,mid})/dr = v_{r,d} v_{\rm K,mid} /2$.  Then, from equation
(\ref{eq:motion_th}), we have
\begin{equation}
v_{\theta,d} - v_{\theta,g} = - \frac{1}{2} T_{s} v_{r,d} \ .
\label{eq:delta_vt}
\end{equation}
Using equation (\ref{eq:veldiff}) and neglecting terms of order
$(h_g/r)^4$ and higher, equation (\ref{eq:motion_r}) is reduced to
\begin{eqnarray}
\frac{d v_{r,d}}{dt} = -\eta \frac{v_{\rm K,mid}^2}{r} + \frac{2 v_{\rm
K,mid}}{r} ( v_{\theta,d} - v_{\theta,g} ) \nonumber \\
- \frac{\Omega_{\rm K,mid}}{T_s} ( v_{r,d} - v_{r,g} ) \ .
\label{eq:motion_r_approx}
\end{eqnarray}
The left hand side is order of $v_{r,d}^2/r$ and is neglected if
$v_{r,d} \ll c$.  Substituting equation (\ref{eq:delta_vt}) into
equation (\ref{eq:motion_r_approx}), we find the radial velocity of
the particle to be
\begin{equation}
v_{r,d} = \frac{T_s^{-1} v_{r,g} - \eta v_{\rm K,mid}}{T_{s} +
T_{s}^{-1}} \ .
\label{eq:vrdrag}
\end{equation}
In the above derivation of the particle's radial velocity, we assume
that in the $z$-direction the particle sedimentation is balanced by the
turbulent diffusion, but that in the $r$-direction the turbulent
effects are neglected.

For particles well coupled with the gas ($T_s \ll 1$), the radial velocity
reduces to
\begin{equation}
v_{r,d} = v_{r,g} + v_{r,{\rm drift}} \ ,
\label{eq:vrd}
\end{equation}
where $v_{r,{\rm drift}} = - \eta T_s v_{\rm K,mid} $ is the relative
velocity from the gas.
Using equations (\ref{eq:powerlaw1}), (\ref{eq:eta}), and
(\ref{eq:stoptime}), 
\begin{eqnarray}
v_{r,{\rm drift}}  = 2 \pi \left( \frac{h_0}{{\rm AU}} \right)^2 \left(
p+ q+ \frac{q+3}{2} \frac{z^2}{h_g^2} \right) \nonumber \\
\times \ r_{\rm AU}^{q+1/2}
T_{s,{\rm mid}} \exp \left( \frac{z^2}{2 h_g^2} \right) \ {\rm AU \
yr}^{-1} \ , 
\label{eq:vrdragap}
\end{eqnarray}
where $T_{s,{\rm mid}}$ is the non-dimensional stopping time at the
midplane.  The radial drift velocity increases exponentially with the
height.  Because the particles are strongly coupled with the gas, the gas drag
force suppresses their drift velocity.  At the midplane where the gas
density is highest, the suppression of the drift is most effective.
The drift velocity at the midplane is
\begin{equation}
v_{r,{\rm drift,mid}} = \sqrt{\frac{\pi^3}{2}} \frac{h_0 s}{\rm AU^2}
\frac{\rho_p}{\rho_0} (p+q) r_{\rm AU}^{-p+q/2-1} \ {\rm AU \ yr^{-1}} \ ,
\label{eq:vrdriftmid}
\end{equation}
which agrees with the derivations of Adachi et al. (1976) and Weidenschilling
(1977).

For large particles decoupled from the gas ($T_s \gg 1$), the radial
motion of the gas does not affect the particles' velocity, so that
$v_{r,d} = -\eta T_s^{-1} v_{\rm K,mid}$.  In this paper, we do not
consider such large particles.

\begin{figure}
\epsscale{1.0}
\plotone{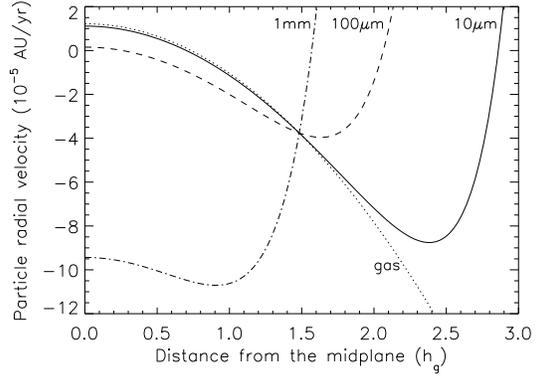}
\caption{
Radial velocity $v_{r,d}$ of dust particles of $s=10 \ \micron$ (solid
line), $100 \ \micron$ (dashed line), and $1 \ {\rm mm}$ (dot-dashed
line) at $10 \ {\rm AU}$. The distance from the midplane $z$ is
normalized by the disk scale height $h_g=0.59 \ {\rm AU}$. The dotted
line shows the radial velocity of the gas $v_{r,g}$.
\label{fig:vrd}
}
\end{figure}

Figure \ref{fig:vrd} shows the radial velocity $v_{r,d}$ of particles
of $s=10 \ \micron$, $100 \ \micron$, and $1 \ {\rm mm}$ at $10 \ {\rm
AU}$ in the standard model disk.  Near the midplane, the drift velocity
$v_{\rm drift}$ (relative to the gas velocity shown by the dotted line) of
$10 \ \micron$ particles is small.  The particles move outward almost
together with the gas.  As the altitude increases and the gas velocity
decreases, the radial velocity of the particles also decreases, then
becomes negative.  At $z \approx 1.5 h_g$ where $\eta$ changes its
sign, the radial pressure gradient vanishes and the gas rotates with the
Keplerian velocity.  The particles co-rotate with the gas at that
location, and have the same radial velocity as the gas.  The drift
velocity changes its sign to become positive.  At high altitude ($z
\ga 2 h_g$) where the gas drag is weak, the particles' drift velocity
begins to increase rapidly, and then the radial velocity becomes
positive again at $z \ga 3 h_g$.  The behavior of the radial velocity
of $100 \ \micron$ particles is qualitatively similar to that of $10 \
\micron$ particles, i.e., it is outward at the midplane, inward
at intermediate altitude, and outward again at high altitude.  The
coupling of $100 \ \micron$ particles to the gas is weaker, so that their
radial velocity is much different from the gas velocity even at the
midplane.  However, the drift velocity at the midplane is still
slightly smaller than the gas outflow velocity and the particles move
outward.  Particles of $1 \ {\rm mm}$ move inward even at the
midplane, because the inward drift velocity of such large particles is
larger than the outflow velocity of the gas.  These large particles also
move outward at high altitude where the gas rotates faster than the
particles.

\subsection{Vertical Distribution of Particles \label{sec:vdist}}

At the beginning stage of star formation from molecular
cloud cores, dust particles in circumstellar disks 
may be well mixed with the gas and the dust-gas ratio may be
uniform throughout the disk.
After the termination of gas infall onto the disks, the disks become to be
hydrostatic and particles begin to sediment toward the midplane because
of the $z$-component of the stellar gravity.
The particles reach the terminal velocity, where the
gravity and the gas drag balance with each other,
\begin{equation}
v_{z,d} = - \Omega_{\rm K,mid} T_{s} z \ .
\label{eq:termnalvel}
\end{equation}
The time scale of the sedimentation is $t_{\rm sed} \sim z/v_{z,d}
\sim T_s^{-1} \Omega_{\rm K}^{-1}$.
If this time scale is much smaller than the radial migration time scale,
$t_r \sim r/v_{r,d} \sim (\alpha + T_s)^{-1} (h_g/r)^{-2} \Omega_{\rm
K}^{-1}$, i.e., if $T_s / \alpha \gg (h_g/r)^2$, then the particles sediment
before the large migration in the $r$-direction.
(At high altitudes or around the midplane, particles may drift to the
opposite direction to the gas, as seen in Fig. \ref{fig:vrd}. Some
particles with $T_s \sim \alpha$ may have the radial migration time
scale $t_r$ much larger than the above estimate. Even for such particles,
the condition $T_s / \alpha \gg (h_g/r)^2$ for the fast sedimentation is
still valid.)
For example, at 10 AU in a disk with $\alpha=10^{-3}$ and $(h_g/r)^2
\sim 10^{-3}$, particles larger than $0.2 \ \micron$ sediment to the
midplane without large radial movement.

We consider disks at later stages, such as the T Tauri stage.
The disks are still turbulent, and
the turbulent motion of the gas stirs dust particles up to high altitude
to prevent dust sedimentation.
An equilibrium distribution of particles in the vertical direction is
achieved by the balance between the sedimentation and the diffusion due
to the turbulent gas.

The turbulent diffusion is modeled in an analogy of molecular diffusion,
provided that the particles are considered to be the passive tracers of fluid,
i.e., if the particles have no influence on the gas motion and have the
similar velocity of the surrounding gas [see e.g., chapter 10 in Monin
\& Yaglom (1971) and section 3.5.1 in Morfill (1985)].
The equation of continuity is written as
\begin{equation}
\frac{\partial}{\partial t} \rho_d + \nabla {\bf \cdot} (\rho_d
\mbox{\boldmath $v$}_d + \mbox{\boldmath $j$} ) = 0 \ ,
\label{eq:dustcont}
\end{equation}
where $\rho_d$ is the particle density, and $\mbox{\boldmath $v$}_d$ is
the particle velocity.
The diffusive mass flux $\mbox{\boldmath $j$}$ is estimated by
\begin{equation}
\mbox{\boldmath $j$} = -\frac{\rho_g \nu}{\rm Sc} \nabla \left(
\frac{\rho_d}{\rho_g} \right) \ ,
\label{eq:difflux}
\end{equation}
where the Schmidt number ${\rm Sc}$ represents the strength of
coupling between the particles and the gas.  For small particles, ${\rm
Sc}$ approaches unity and the particles have the same diffusivity as the
gas, while it becomes infinite for large particles.
For intermediate particle sizes, ${\rm Sc}$ can be as small as $0.1$,
which means that particle diffusion occurs effectively (see e.g.,
Fig. 1 in Cuzzi et al. 1993).  In our standard model, we use ${\rm
Sc}=1$.
If ${\rm Sc}$ is not unity, the particles are not the passive tracers
which completely follow the fluid motion. In this case, the formulation
according to the analogy of molecular diffusion may be inappropriate. In
addition, if the velocity of sedimentation, $v_{z,d}$, is comparable to
or larger than the turbulent velocity, the particle could not be the
passive tracers. The estimate of the diffusive mass flux by equation
(\ref{eq:difflux}) should be considered just as a ``gradient diffusion
hypothesis.''

In a steady state of axisymmetrical disks, $\partial/ \partial t$ and
$\partial/ \partial \theta$ in equation
(\ref{eq:dustcont}) is zero. In addition, for particles satisfying the
condition $T_s / \alpha \gg (h_g/r)^2$ (i.e., for particles sedimenting
fast without large radial migration), we see that $\partial (\rho_d
v_{z,d}) / \partial z \gg \partial (\rho_d v_{r,d}) / \partial r$.
Therefore, the mass flux in the $z$-direction must be zero, i.e., 
\begin{equation}
\rho_d v_{z,d} - \rho_g \frac{\nu}{\rm Sc} \frac{\partial}{\partial z} \left(
\frac{\rho_d}{\rho_g} \right) = 0 \ .
\label{eq:dustdiff}
\end{equation}
This equation is the same as the one derived by Dubrulle, Morfill, \&
Sterzik (1995).
Solving equation (\ref{eq:dustdiff}) with equations
(\ref{eq:gasdensity}), (\ref{eq:hg}), (\ref{eq:nu}),
(\ref{eq:stoptime}), and (\ref{eq:termnalvel}) gives the particle density
\begin{equation}
\rho_{d}(r,z) = \rho_{d}(r,0) \exp \left[ - \frac{z^2}{2h_g^2} -
\frac{{\rm Sc} T_{s, {\rm mid}}}{\alpha}
\left( \exp \frac{z^2}{2 h_g^2} - 1 \right) \right] \ .
\label{eq:dustdistr}
\end{equation}
The first term in the exponential comes from the gas distribution and
the second term represents the sedimentation.
In the limit of tight coupling of the particles and the gas
($T_{s, {\rm mid}}=0$), the particle
distribution is the same as that of the gas.  The surface mass density
of dust particles is
\begin{equation}
\Sigma_d (r) = \int^{+\infty}_{-\infty} \rho_{d} (r,z) dz
\ .
\label{eq:dustsurface2}
\end{equation}

\begin{figure}
\epsscale{1.0}
\plotone{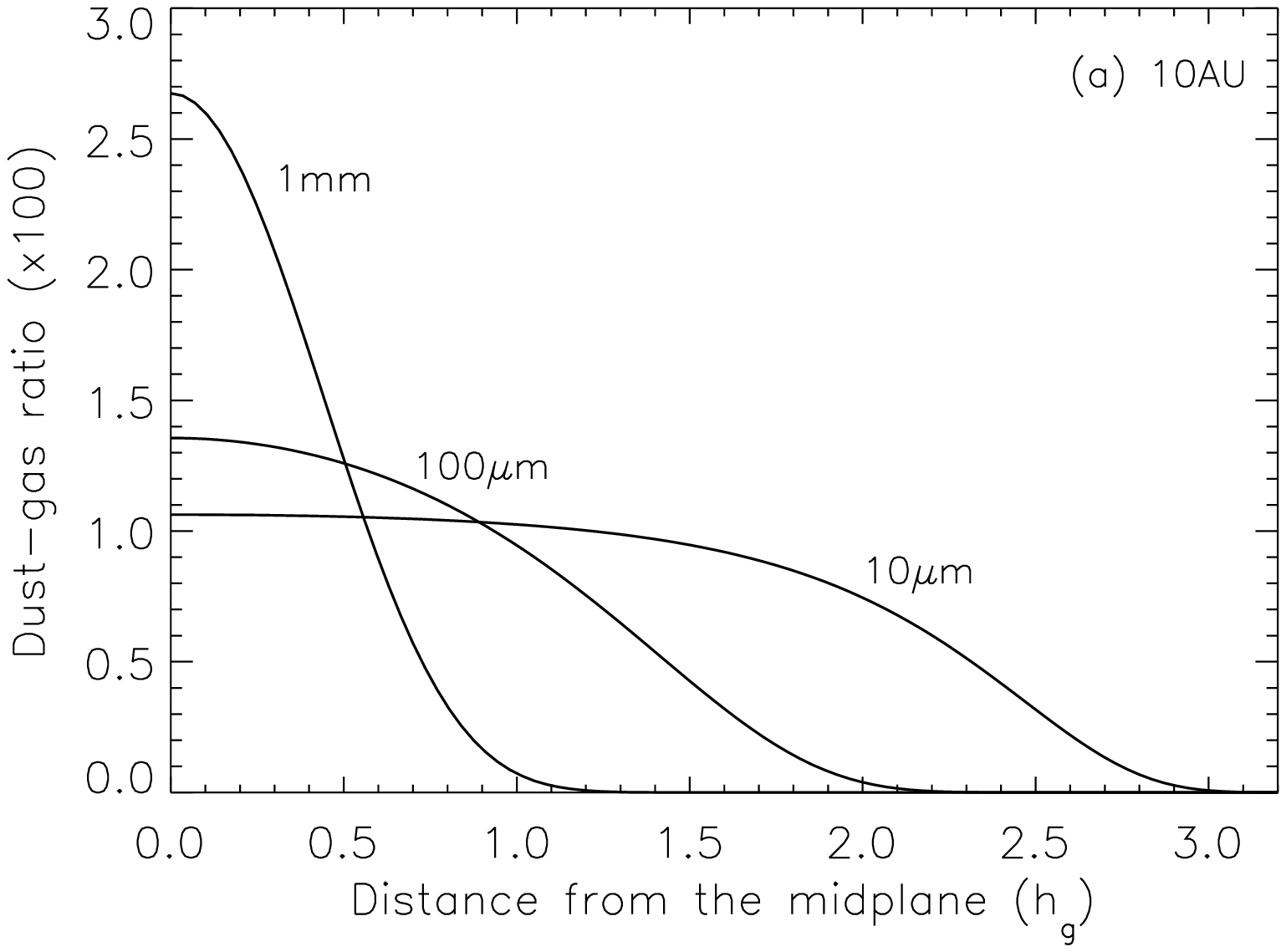}
\plotone{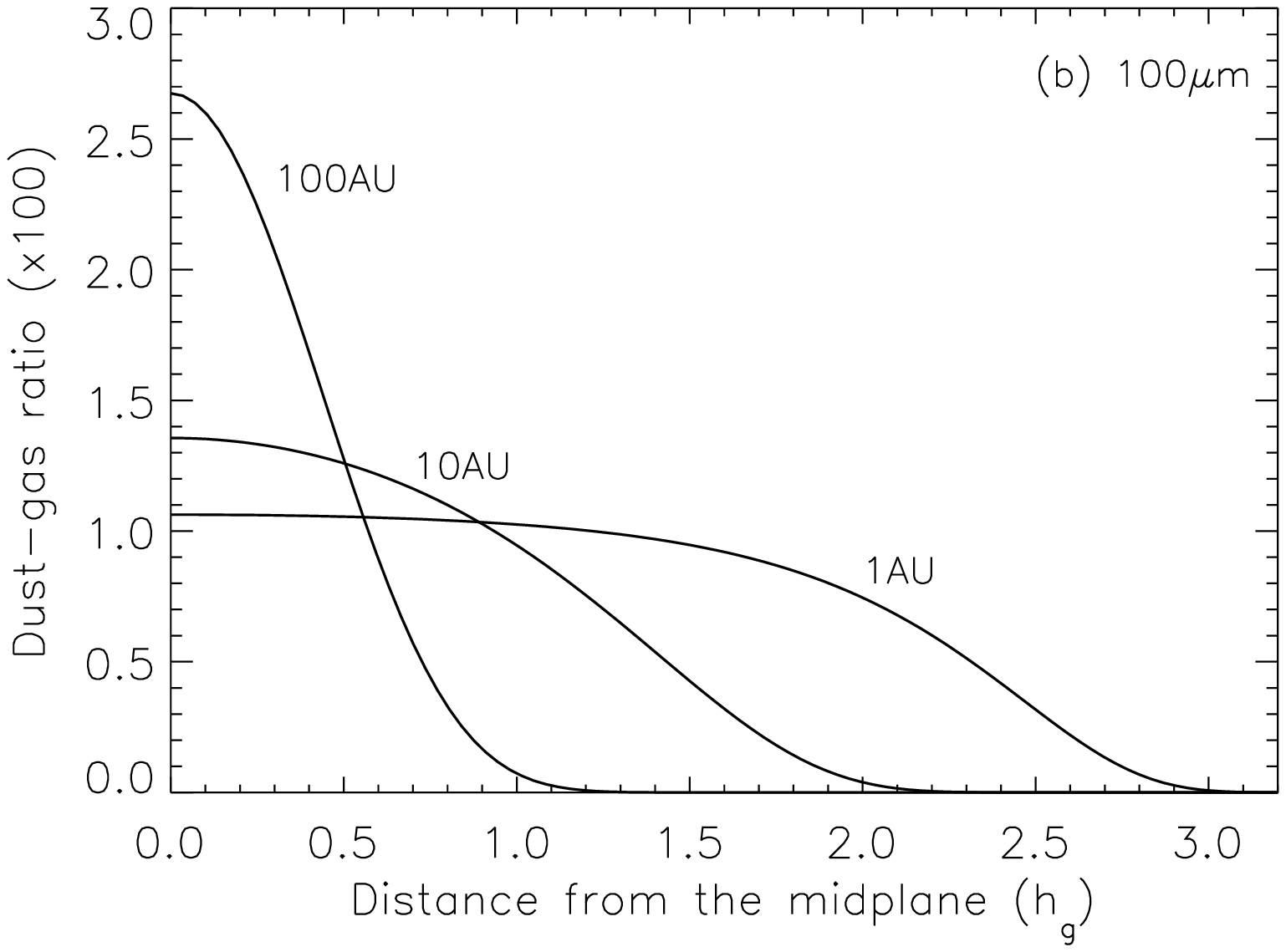}
\caption{
Dust-gas ratio, $\rho_d / \rho_g$, against the distance from the midplane
($a$) for various sizes at $10$ AU and ($b$) at various locations for
$100 \ \micron$. The distance from the midplane $z$ is normalized by the
disk scale height $h_g$. The averaged dust-gas ratio is assumed to be
$0.01$. ($a$) Three solid lines show the dust-gas ratio of particles of
$1$ mm, $100 \ \micron$, and $10 \ \micron$ from the left line.
($b$) Three solid lines show the dust-gas ratio at $100$ AU,
$10$ AU, and $1$ AU from the left line.
\label{fig:verden}
}
\end{figure}

Figure \ref{fig:verden}$a$ shows variation of the dust-gas ratio in
the vertical direction at $10 \ {\rm AU}$ in the standard model disk.
We see the sedimentation of particles.
\footnote{ We obtained slightly thinner distribution of particles than
that by Dubrulle et al. (1995) which is shown in their Figure 3. They
assumed that the stopping time of particles, which is inversely
proportional to the gas density, to be constant and used the value at
the midplane. This assumption causes the dust disk to puff out.}
Large particles ($\ga 1 \ {\rm mm}$) sediment around the midplane,
while small particles ($\la 10 \ \micron$) spread over a few scale
height of the gas disk.  These small particles are stirred up to high
altitude by the turbulence of the gas.  The gas drag force becomes weaker
at higher altitude where the gas density decreases, so the gas cannot
sustain the dust particles there.  Thus, the particle density drops
rapidly at some altitude and the dust disk obtains a relatively sharp
surface (e.g., at $z \sim 2.5h_g$ for $10 \ \micron$ particles).  The
sedimentation of particles is more effective in the outer part of the
disk, farther away from the star (see Fig. \ref{fig:verden}$b$).  For
example, $100 \ \micron$ particles at $1 \ {\rm AU}$ spread over the
gas disk, while the particles at $100 \ {\rm AU}$ concentrate around
the midplane.  At the outer part of the disk, the gas density is lower, so
that the gas drag is weaker and turbulence cannot loft the particles
to high altitudes.

Dust particles are well mixed with the gas if
\begin{equation}
\frac{{\rm Sc} T_{s, {\rm mid}}}{\alpha}
\left( \exp \frac{z^2}{2 h_g^2} - 1 \right) \ll 1 \ ,
\end{equation}
i.e., below the height
\begin{equation}
\frac{z}{h_g} = \left[ 2 \log \left( \frac{\alpha}{{\rm Sc} T_{s, {\rm
mid}}} + 1 \right) \right]^{1/2} \ .
\end{equation}
In the standard model, particles smaller than
\begin{equation}
s \ll 20 \ r_{\rm AU}^{-1} \ \micron
\label{eq:condmix}
\end{equation}
uniformly distribute with regard to the gas under the height $z=3 h_g$.

\subsection{Net Radial Velocity}

As discussed in \S\ref{sec:vrd}, the radial velocity of particles
varies with the altitude.  In this subsection, we discuss the net radial
velocity averaged in the vertical direction.

\subsubsection{Particles Well Mixed with the Gas}

At the beginning stage of formation of circumstellar disks, small
particles would be distributed uniformly in the disk gas. Before the
particles have undergone enough sedimentation, their net radial velocity
averaged in the vertical direction is
\begin{equation}
\langle v_{r,d} \rangle = \frac{1}{\Sigma_g} \int_{- \infty}^{+ \infty}
\rho_g v_{r,d} dz \ .
\label{eq:netvelmix}
\end{equation}
Figure \ref{fig:netvrd1} shows the net radial velocities of small particles.
The velocity of particles approaches the gas velocity as the
distance from the star becomes smaller.  Near the star, the gas
density is high enough to induce almost all particles to move together
with the gas.  As the distance from the star increases, the gas density
decreases and particles at high altitude begin to drift outward from
the gas.  At sufficiently large distance from the star, the net radial
velocity can be positive, i.e., particles move outward.
However, before they move over a large distance in the radial
direction, they also sediment to the midplane.  The time scale of the
outward radial motion is $t_{\rm out} \sim r/ \langle v_{r,d} \rangle
\sim 1/ (\eta T_s \Omega_{\rm K})$ (for outflowing particles,
$|v_{r,g}|$ is smaller than $|v_{r, {\rm drift}}|$ and is neglected in
the estimate of $t_{\rm out}$), while the sedimentation time scale 
is $t_{\rm sed} \sim z/ v_{z,d} \sim 1/ (T_s \Omega_{\rm K})$.  Because
$\eta \ll 1$, $t_{\rm out} \gg t_{\rm sed}$.  The outward motion of
those particles which are well mixed with the gas may proceed shortly
after the formation of circumstellar disks, but it does not
significantly modify the radial distribution of dust particles.  The
evolution of radial distribution occurs primarily through the motion
of sedimented particles.

\subsubsection{Removal of Outflow at High Altitude by the Sedimentation}

The net radial velocity of sedimented particles is
\begin{equation}
\langle v_{r,d} \rangle = \frac{1}{\Sigma_d} \int_{- \infty}^{+ \infty}
\rho_d  v_{r,d} dz \ .
\label{eq:netvrd}
\end{equation}
The function $\rho_d v_{r,d} / \Sigma_d$, which is
proportional to the mass flux, is shown in Figure \ref{fig:fvrd}.
Because of the sedimentation to the midplane, which is efficient for
larger particles, the mass flux is dominated by the particles near the
midplane.  For example, the mass flux of $10 \ \micron$ particles is
mainly carried by particles moving outward around the midplane ($|z|
\la 0.7h_g$) and by those moving inward at intermediate altitude
($0.7h_g \la |z| \la 2.9 h_g$).  Although particles at high altitude
($|z| \ga 2.9 h_g$) move outward, their contribution to the mass flux
is negligibly small.  Very large particles, for example $1 \ {\rm mm}$
particles, move inward even at the midplane.  Thus, the sedimentation
causes the vast majority of particles to move inward.

Sedimentation effectively removes outflowing particles at high
altitude.  This flow pattern is seen as follows.  The particles
flowing outward have a drift velocity larger than the gas inflow velocity,
i.e., $|v_{r,{\rm drift}}| > |v_{r,g}|$.  From equations
(\ref{eq:vrg}) and (\ref{eq:vrdragap}), it is seen that $\alpha^{-1}
T_{s,{\rm mid}} \exp (z^2/2 h_g^2) \ga 1$ for such particles.
However, from equation (\ref{eq:dustdistr}), we also see that the
density of these particles is as small as $\rho_d (z)/
\rho_d (0) \la e^{-1}$.  Therefore, very few particles 
remain in the outflow region at high altitude.

If the sedimentation is so effective that most particles concentrate
around the midplane ($|z| \la h_g$), e.g., for particles larger
than $1 \ {\rm mm}$ at $10 \ {\rm AU}$ (see Fig. \ref{fig:verden}$a$),
the inward drift velocity at the midplane is larger than the gas outflow
velocity.  Thus, such particles flow inward.  Again from equation
(\ref{eq:dustdistr}), we see that $T_{s,{\rm mid}} / \alpha \ga 1$ for
such particles.  Then, it is seen from equations (\ref{eq:vrg}) and
(\ref{eq:vrdragap}) that $|v_{r,{\rm drift}}| \ga
|v_{r,g}|$ at the midplane. 
This inequality implies that when particles grow up sufficiently
large sizes to sediment toward and to concentrate around the midplane,
they become decoupled from the outflow motion of the gas.  If the
sedimentation is so efficient that all particles concentrate in a thin
layer at the midplane ($T_{s,{\rm mid}} / \alpha \gg 1$), the radial
motion of the particles would not affected by the radial motion
of the gas.  In this limit, the particles' motion is similar to that
deduced by Adachi et al. (1976) and Weidenschilling (1977).

\subsubsection{Net Velocity of Sedimented Particles}

Figure \ref{fig:netvrd2}$a$ shows the net radial velocities of sedimented
particles.  These net radial velocities are negative for particles of all
sizes.  Particles rapidly move inward when they are at large distances
from the star.  In the outer regions of the disk, particles sediment
and concentrate at the midplane, and their net radial velocity, which is much
faster than the gas outflow velocity $v_{r,g}$, is approximated by the
drift velocity at the midplane (eq. [\ref{eq:vrdriftmid}]).  Their
inward velocity decreases as they approach the star, because the
suppression of inward velocity by the gas drag becomes stronger.  The
time scale of their orbital decay is $r / \langle v_{r,d}
\rangle \propto r^{2+p-q/2}$.  When they approach the location where
the gas density is dense enough to make $T_{s,{\rm mid}}/\alpha \la
1$, the inward drift velocity becomes smaller than the gas outflow
velocity at the midplane.  The particles begin to move with the gas
flow.  At the same time, particles are spreaded over more than the
disk scale height.  The particles' net radial velocity approaches the
net gas velocity (the dashed line in Fig. \ref{fig:netvrd2}$a$).  However,
because the particle distribution concentrates slightly more to the
midplane than the gas distribution, the number of particles riding on
the outflowing gas around the midplane is larger than in the case
where the particle distribution is same as the gas.  The
enhancement of outflowing particles retards their net inflow velocity
to values below that of the gas.  At the inner part of the disk, the
net inward velocity of particles is always slower than the gas
velocity.  Near the star, the particles mix more with the gas and
their motion converges with that of the gas. The particles' inward
velocity has a minimum magnitude, $\langle v_{r,d} \rangle = 5.2
\times 10^{-6} \ {\rm AU/yr}$, which is about half of the gas
velocity.  Particles smaller than $10 \ \micron$ have a slower inward
velocity than the gas throughout nearly the entire disk ($r \la 100 \
{\rm AU}$),
while particles larger than $1 \ {\rm mm}$ have slower velocity only
in the innermost part of the disk ($r \la 1 \ {\rm AU}$).  The
difference in the inward velocity induces the size fractionation of
particles, as discussed in \S\ref{sec:steady_dist} below.  The dust-gas
ratio may increase during the viscous evolution of disks, because the
accretion velocity of small particles is slower than that of the gas.

\begin{figure}
\epsscale{1.0}
\plotone{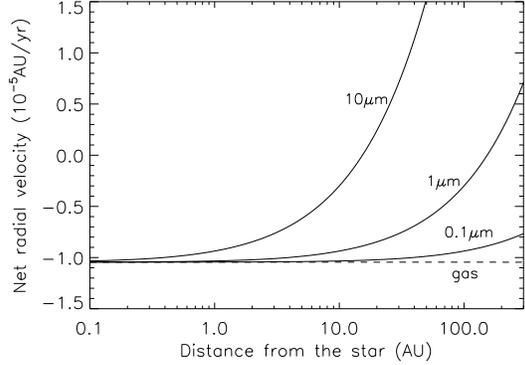}
\caption{
Net radial velocity of particles well mixed with the gas, $\langle v_{r,d}
\rangle$. The three solid lines correspond to $s=10 \ \micron$, $1 \
\micron$, and $0.1 \ \micron$ particles from the upper line.
The dashed line shows the gas accretion velocity.
\label{fig:netvrd1}
}
\end{figure}

\begin{figure}
\epsscale{1.0}
\plotone{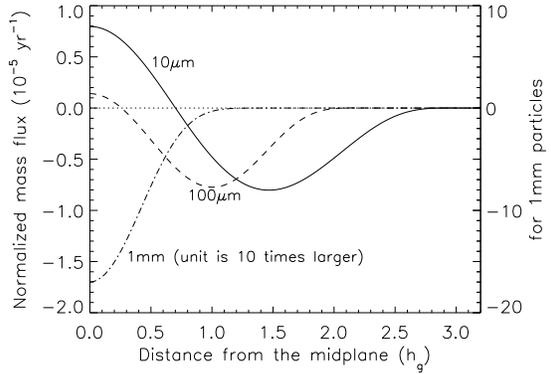}
\caption{
Normalized mass flux $\rho_d v_{r,d} / \Sigma_d$ at $10$
AU. The solid, dashed, and dot-dashed lines are for $s=10 \ \micron$,
$100 \ \micron$, and $1$ mm particles, respectively. For $1$ mm
particles (dot-dashed line), $1/10$ of the value is plotted (or refer
the ordinate at the right hand side). The distance
from the midplane $z$ is normalized by the disk scale height $h_g=0.59$
AU. The dotted line shows the zero velocity for reference.
\label{fig:fvrd}
}
\end{figure}

\begin{figure}
\epsscale{1.0}
\plotone{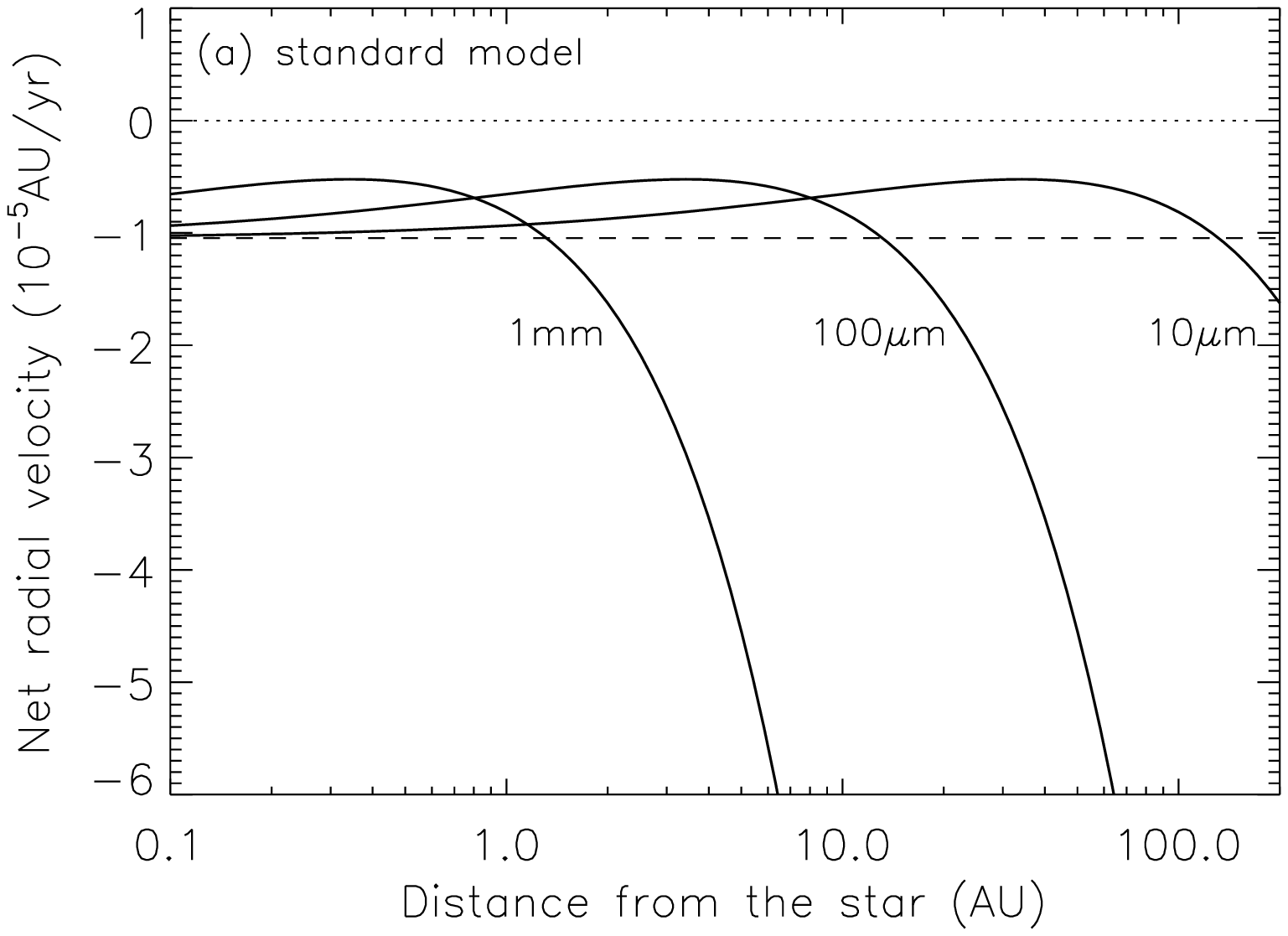}
\plotone{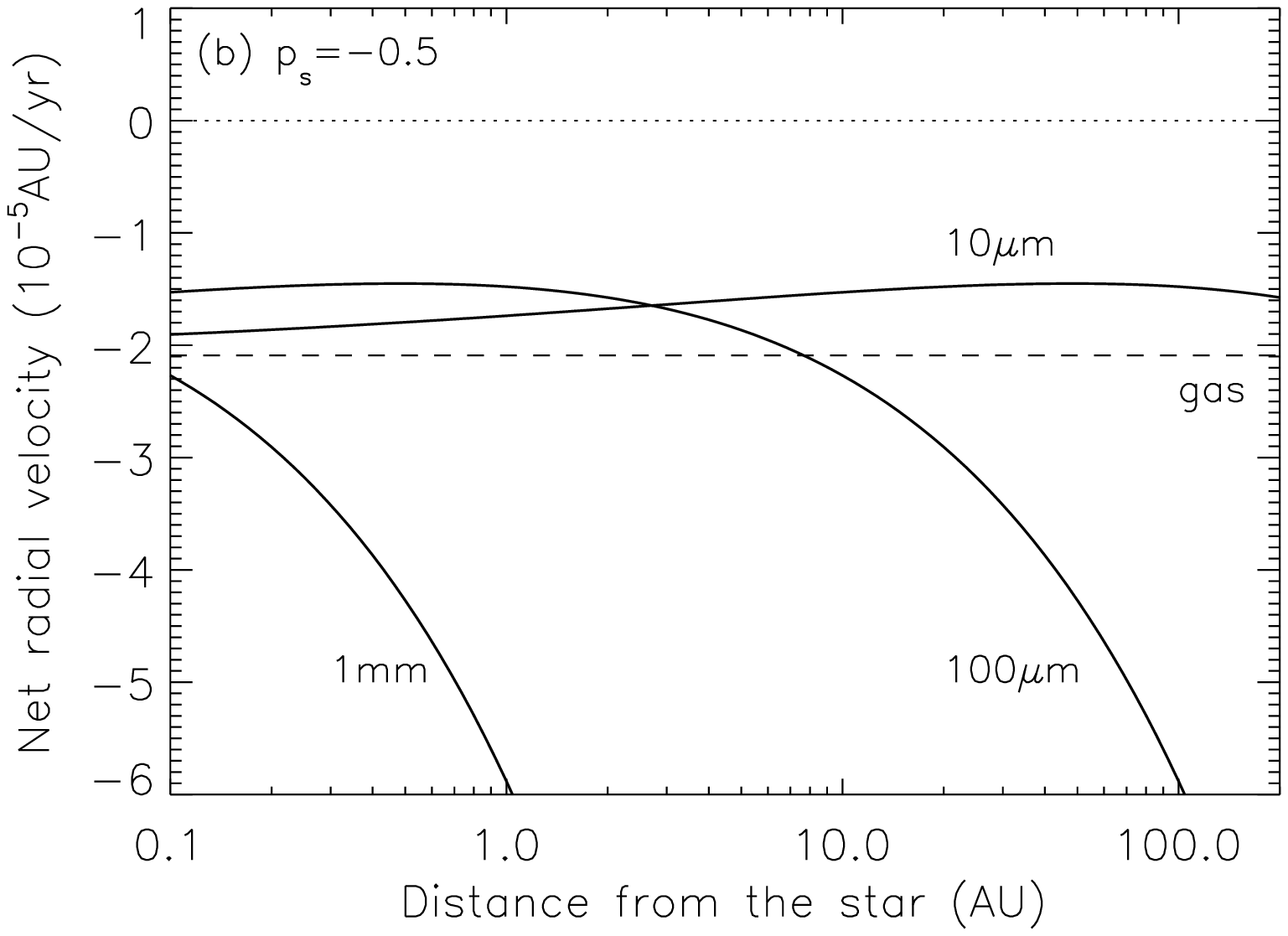}
\caption{
Net radial velocity $\langle v_{r,d} \rangle$ of sedimented particles.
Three solid lines correspond to $s=1$ mm, $100 \ \micron$, and $10 \
\micron$ particles from the left line.
The accretion velocity of the gas is shown by the dashed line, and the
dotted line shows the zero velocity for reference.
($a$) the standard model ($p_s=-1.0$, $q=-0.5$, and ${\rm Sc}=1.0$)
($b$) $p_s=-0.5$.
($c$) $p_s=-1.3$.
($d$) ${\rm Sc} = 10$.
\label{fig:netvrd2}
}
\end{figure}

Small particles which satisfy the condition (\ref{eq:condmix}) are mixed
well with the gas. The net radial velocities of such small particles may
be calculated by equation (\ref{eq:netvelmix}), even after larger
particles have undergone the sedimentation. It is seen from
comparison between Figures \ref{fig:netvrd1} and \ref{fig:netvrd2}$a$
that, e.g., for $10 \ \micron$ particles, the radial velocities are
about the same in both figures for $r \la 1$ AU, while they deviate
significantly between these figures for $r \ga 10$ AU.
As shown in Figure \ref{fig:netvrd1}, the radial velocity of $10 \
\micron$ particles at $r \ga 10$ AU is outward when the particles are
mixed with the gas, i.e., in the very first stage of disk formation.
However, after the particle sedimentation, the radial velocity becomes
inward even at $r \ga 10$ AU, as shown in Figure \ref{fig:netvrd2}$a$.
It is concluded that the particle sedimentation suppresses the outward
flow of particles in the standard model.

\subsubsection{Various Models}

In Figure \ref{fig:netvrd2}$b$, we show the case with $p_s = -0.5$
where the surface density profile of the gas is flatter than in the
standard model.  The velocity profiles are qualitatively similar to
those in the standard model.  Particles of $10 \ \micron$ move slower
than the gas throughout nearly the entire disk, while $1 \ {\rm mm}$
particles rapidly migrate inward.

In Figure \ref{fig:netvrd2}$c$, the case with a steeper surface density
gradient is shown.  In this model, we adopt $p_s = -1.3$ which
satisfies the condition (\ref{eq:condaccretion}) for inward gas
accretion flow.  In this case, we see that the net radial velocity of
particles becomes positive at some locations.  In these regions, the
accretion of particles is prevented.  The particles flowing inward
from large distances terminate their migration at the location where
the net radial velocity becomes zero, and then accumulate at that particular
orbital radius.  Particles located just inside this critical radius flow
outward and 
accumulate there also.  Particles at the innermost part of the disk
flow inward onto the star.  Size fractionation occurs because the
location of the accumulation depends on the particle size.  The
accumulation continues as particles drift radially inward from large
radii until particles in the outer disk are significantly depleted or the
number density of accumulated particles becomes so high that the removal
of particles through 
coagulation or collisional destruction becomes efficient.  Although
accretion of dust particles terminates at some locations, the gas disk
continues to accrete onto the star.  The dust-gas ratio increases as
the accretion of the gas proceeds.

\subsubsection{Self-similarity of the Velocity Profile}

The functions of net radial velocities shown in Figure \ref{fig:netvrd2} for
various sizes have self-similar forms. From equations (\ref{eq:vrg}),
(\ref{eq:vrd}), (\ref{eq:vrdragap}), (\ref{eq:dustdistr}), 
(\ref{eq:dustsurface2}), and (\ref{eq:netvrd}), the net radial velocity can be
written as
\begin{equation}
\langle v_{r,d} \rangle  = 2 \pi \alpha \left( \frac{h_0}{\rm AU}
\right)^2 r_{\rm AU}^{q+1/2} F \left( p, q, {\rm Sc} ; \frac{T_{s,{\rm
mid}}}{\alpha} \right) \ {\rm AU \ yr^{-1}} \ ,
\label{eq:netvelapprx}
\end{equation}
where $F$ is a function of $p$, $q$, ${\rm Sc}$, and $T_{s,{\rm mid}}/\alpha$, 
\begin{eqnarray}
\lefteqn{F \left( p,q,{\rm Sc} ; \frac{T_{s,{\rm mid}}}{\alpha} \right)
= }
\nonumber \\
&  \int_{- \infty}^{+ \infty} \left\{ -3p -2q -6 - (5q+9) z^{\prime
2} \right. \nonumber \\
& + \left. \left[ p+q+ (q+3) z^{\prime 2} \right]
\frac{T_{s,{\rm mid}}}{\alpha} \exp z^{\prime 2} \right\}
\nonumber \\
& \times \exp \left[ -z^{\prime 2} - \frac{{\rm Sc} T_{s,{\rm
mid}}}{\alpha} \left( \exp z^{\prime 2} -1 \right) \right]
dz^{\prime} \nonumber \\
& \times \left\{ \int_{- \infty}^{+ \infty}  \exp
\left[ -z^{\prime 2} - \frac{{\rm Sc} T_{s,{\rm mid}}}{\alpha} \left( \exp
z^{\prime 2} -1 \right) \right] dz^{\prime} \right\}^{-1} \ ,
\end{eqnarray}
and $z^{\prime} = z/\sqrt{2} h_g$. If we vary the properties of
particles, keeping $T_{s,{\rm mid}}/\alpha$ to be constant,
the function $F$ yields the same value.  For example, when we vary the
particle size $s \rightarrow s^{\prime}$, then transform the orbital
radius as $r \rightarrow r^{\prime} = (s^{\prime} /s)^{1/[p +(q+3)/2]}
r$, we have the same value of $T_{s,{\rm mid}}/\alpha$ and the same
functional form of $F$.  In our models, $q=-1/2$, so $\langle
v_{r,d} \rangle$ depends on $r$ only through the function $F$.  In
this case, the profiles of the net radial velocities for different sizes are
obtained by the transformation in the $r$-direction from one profile.
In Figure \ref{fig:f}, the function $F$ is plotted for various values of
the Schmidt number ${\rm Sc}$.

\subsubsection{Various Schmidt Numbers}

In the standard model, the net radial velocity of particles is inward
everywhere in the disk.  This flow pattern is partially due to the
removal of particles, through sedimentation, from high altitudes where
they would flow outward.  For particles which are sufficiently large
to concentrate near the midplane, the magnitude of their inward drift
velocity is larger than that of the outflowing gas. 
Thus, the strength of the sedimentation is always moderate enough to
remove outflowing particles. The sedimentation is controlled by
coupling of particles to the 
turbulent motion of the gas.  If coupling is relatively strong (${\rm
Sc}$ is smaller than unity), the sedimentation would be less effective
and particles would mix more completely to the gas, while if it is relatively
weak (${\rm Sc}$ is larger), particles would concentrate more to the 
midplane.  Usually ${\rm Sc}$ is assumed to be unity, but the actual
value may be different from unity.  Some experimental data are plotted
in Figure 1 in Cuzzi et al. (1993) which show ${\rm Sc}$ as small as
$0.1$ when the Stokes number ${\rm St}$ is about $0.01$ (the Stokes
number is considered to be the same order of the non-dimensional
stopping time $T_s$ in our models).
For large particles (${\rm St} \sim T_s \ga 1$), ${\rm Sc}$ increases
with the particle size.

Figure \ref{fig:netvrd2}$d$ shows the net radial velocities in the case where
the sedimentation is efficient (${\rm Sc}=10$).
The net radial velocity becomes outward at its peak.
Thus, an accumulation of particles and an increase in the dust-gas ratio
would occur.
Because of the strong sedimentation, the dust disk is much thinner than
in the standard model (${\rm Sc}=1$) and concentrates more to  
the midplane where the gas flows outward.
In the outer part of the disk (for example $r \ga 10$ AU for $100 \
\micron$ particles), the inward drift velocity of particles (relative to
the gas) at the midplane is faster than the outward gas velocity.
As the particles drift in, the inward velocity decreases, and
at some distances from the star ($r \sim 10$ AU for $100 \ \micron$
particles), the inward drift velocity at the
midplane becomes smaller than the outward gas velocity, and the
particles around the midplane flow outward.
At such distances, the dust disk is still concentrated around the
midplane, and the net radial velocity is dominated by the particles around
the midplane and is outward.
As the distance from the star decreases, the particles mix more with the
gas and the concentration to the midplane becomes weaker.
The net radial velocity approaches the gas velocity, which is inward, at the
innermost part of the disk.

If the Schmidt number ${\rm Sc}$ is less than unity, the sedimentation
of particles is less efficient. 
Therefore, the particles are always well mixed with the gas at where the
inward drift velocity of particles is smaller than the gas outflow
velocity (i.e., at where the particle outflow at the midplane occurs).
Thus, the net radial velocity of particles is inward everywhere.

Figure \ref{fig:f} shows the function
$F$ for various values of the Schmidt number, ${\rm Sc}=0.1$, $1$, and
$10$.  The net radial velocity has the same sign as $F$ (see eq.
[\ref{eq:netvelapprx}]).
It is seen that the outflow of dust particles occurs if the
sedimentation is efficient (${\rm Sc} \ga 10$).

\section{Discussion and Summary}

\subsection{Steady Density Distribution of Dust Particles
\label{sec:steady_dist}}

Because the inward velocity of particles depends on their size,
particles of different sizes accumulate at different locations.  In
this subsection, we discuss the density distribution of dust particles
as a consequence of their radial flow and show how their size
fractionation may occur.  We adopt the models in which the net
velocity of particles is always inward (${\rm Sc}=1$ and $p_s \ge
-1.0$).  We assume that the distribution of particles approaches
a steady state after a brief stage of initial evolution.  When a
steady state is achieved, the mass flux of dust particles becomes
constant in the radial direction.  We calculate the mass flux of
particles as
\begin{equation}
\frac{d \dot{M_d}}{ds} = 2 \pi r \langle v_{r,d} \rangle \frac{d
\Sigma_d}{ds} \ ,
\label{eq:mflux}
\end{equation}
where $\dot{M_d}$ and $\Sigma_d$ are the mass flux and the surface
density, respectively, of particles smaller than size $s$.
For simplicity, we neglect three physical processes.  First, the
evolution of the particle size through the coagulation, collisional
destruction, condensation, and sublimation is neglected, i.e., we
assume the mass flux $d \dot{M_d}/ds$ for each size range is constant in
the radial direction.  Second, in turbulent disks, the particles' mass
flux comes not only from the mean flow with an average velocity
$\langle v_{r,d} \rangle$ but also from the turbulent diffusion of
particles, which appears as $\mbox{\boldmath $j$}$ in equation
(\ref{eq:dustcont}) for example.
We neglect the mass flux from the turbulent diffusion.
Third, we assume the structure of the 
gas component of the disk is entirely determined by the gas itself.  We
neglect the feedback drag induced by the particles on the gas.  In the
limit that the spatial density of the dust component becomes
comparable to that of the gas near the midplane, this effect would
speed up the azimuthal velocity of the gas to the Keplerian value and
quench the radial migration of the dust (Cuzzi et al. 1993).  Such a
dust concentration requires substantial sedimentation 
of relatively large particles.  Contributions from all of these
factors will be investigated in future grain-evolution
calculations.  Here, we adopt the simplest assumptions to focus on
showing the size fractionation of particles.  The mass flux of
particles in all size range is
\begin{equation}
\dot{M}_{d,{\rm all}} = \int_{s_{\rm min}}^{s_{\rm max}} \frac{d
\dot{M_d}}{ds} ds \ ,
\end{equation}
where $s_{\rm min}$ and $s_{\rm max}$ are the minimum and maximum
sizes of particles, respectively.  The surface density in the steady
state is calculated from equation (\ref{eq:mflux}) with given
$\dot{M_d}$.  Figure \ref{fig:surden}$a$ shows the surface density of
the dust component composed of single size particles.  Particles of
different sizes have different density profiles.  If the dust
component is composed of relatively large particles, it would be
concentrate at the inner region of the disk.  Thus, as particles grow
in size, their surface density distribution becomes more centrally
concentrated.  Figure \ref{fig:surden}$b$ shows the surface density,
assuming the size distribution of particles is a power law with
index $-3.5$, $s_{\rm min}=0.1 \ \micron$, and $s_{\rm max}=10 \
{\rm cm}$.  The surface density distribution of the dust particles is
different from that of the gas.  The power law index of the dust 
distribution $\Sigma_d$ is about $-1.5$ for a gas disk with index $p_s=-1.0$
($-1.2$ for a gas disk with $p_s=-0.5$).  The implied power law index,
$-1.5$, is similar to the value anticipated from the present mass
distribution of planets in the solar system (Hayashi et al. 1985).
However, note that the density distributions in Figure 
\ref{fig:surden} are derived assuming no size evolution of particles,
and no turbulent diffusion in the radial direction.  The growth of
particles during the radial flow adds a source term in the equation of
continuity.  The evolution of the dust density profile should be
investigated further by taking particle growth and turbulent
diffusion into account.

\subsection{Evolution of the Dust-gas Ratio}

The accretion velocity of dust particles is different from that of
the gas.  This difference causes evolution of the dust-gas ratio.  In the
standard model, $10 \ \micron$ particles between $10$ and $100$ AU have
an inflow velocity which is about half of the gas velocity.  Thus, if
the mass of the dust disk is dominated by $10 \ \micron$ particles (for
example, if the particles have size distribution $n \propto s^{-3.5}$
with maximum size $10 \ \micron$), the dust-gas ratio would increase
through the gas accretion.  For example, if the initial mass of the
gas disk is $0.11 M_{\sun}$ and it reduces to $0.01 M_{\sun}$ after
viscous accretion, the dust-gas ratio would increase to be 6 times of
its initial value.  The increase in the dust-gas ratio speeds up the
formation of planetesimals through mutual collisions.  Their enhanced
abundance may also cause the dust component to become unstable and
promote the planetesimal formation through the gravitational
instability (Goldreich \& Ward 1973; Sekiya 1998).

If the particle growth proceeds to make $1$ mm to $1$ cm particles
during the gas accretion phase, such large particles migrate inward
rapidly and accumulate in the inner part of the disk (see Fig
\ref{fig:surden}$a$).  This size fractionation causes an increase in the
dust-gas ratio at the inner disk, while this process may decrease the
dust-gas ratio at the outer disk, resulting in a dust disk concentrated
to the inner part of the gas disk.

\subsection{Summary}

The radial migration of dust particles in accretion disks is studied.
Our results are as follows.

1. Dust particles move radially both inward and outward by the gas
drag force.  Particles at high altitude ($|z| \ga 2 h_g$) move outward
because they rotate slower than the gas whose pressure gradient force is
inward.  Small particles ($s \la 100 \ \micron$ at $10 \ {\rm AU}$)
near the midplane ($|z| \la h_g$) are advected by the gas outflow.  On the
other hand, particles at intermediate altitude and large particles ($s
\ga 1 \ {\rm mm}$ at $10 \ {\rm AU}$) move inward.

2. The net radial velocity, averaged in the vertical direction, is usually
inward, provided that the radial gradient of the gas surface density is
not too steep ($p_s \ga -1.3$).  Sedimentation removes outflowing
particles from high altitudes.  Small particles, which can be
advected by the outflowing gas around the midplane, do not
concentrate at the midplane.  In the inner part of the gas disk ($r
\la 100$ AU for $10 \ \micron$ particles), the inflow velocity of
particles is smaller than the gas accretion velocity, resulting in an
increase in the dust-gas ratio.

3. The particle sedimentation would be efficient if dust-gas coupling
is relatively weak (${\rm Sc} > 1.0$).
If the sedimentation is so efficient (${\rm Sc} \ga 10$),
the number of outflowing particles around the midplane would
be large, and the direction of the net radial velocity of particles
would change to outward at some distances from the star.  Accumulation
of particles at such locations serves to increase the local dust-gas
ratio.

4. The inflow velocity of particles depends on the particle size.
Therefore, the inflow causes the size fractionation of particles.
Larger particles accumulate at distances closer to the star.

\acknowledgements

We are grateful to the anonymous referee who gave us important
criticism, especially on the improper treatment of particle diffusion in
the original version of the manuscript. His/her suggestions also
considerably improved the paper. We thank Greg Laughlin for
careful reading of the manuscript and Jeff Cuzzi for discussions on the
turbulent motion of dust particles.
This work was supported in part by an NSF grant AST 99 87417 and in part
by a special NASA astrophysical theory program that supports a joint
Center for Star Formation Studies at UC Berkeley, NASA-Ames Research
Center, and UC Santa Cruz.  



\clearpage

\begin{figure}
\epsscale{1.0}
\plotone{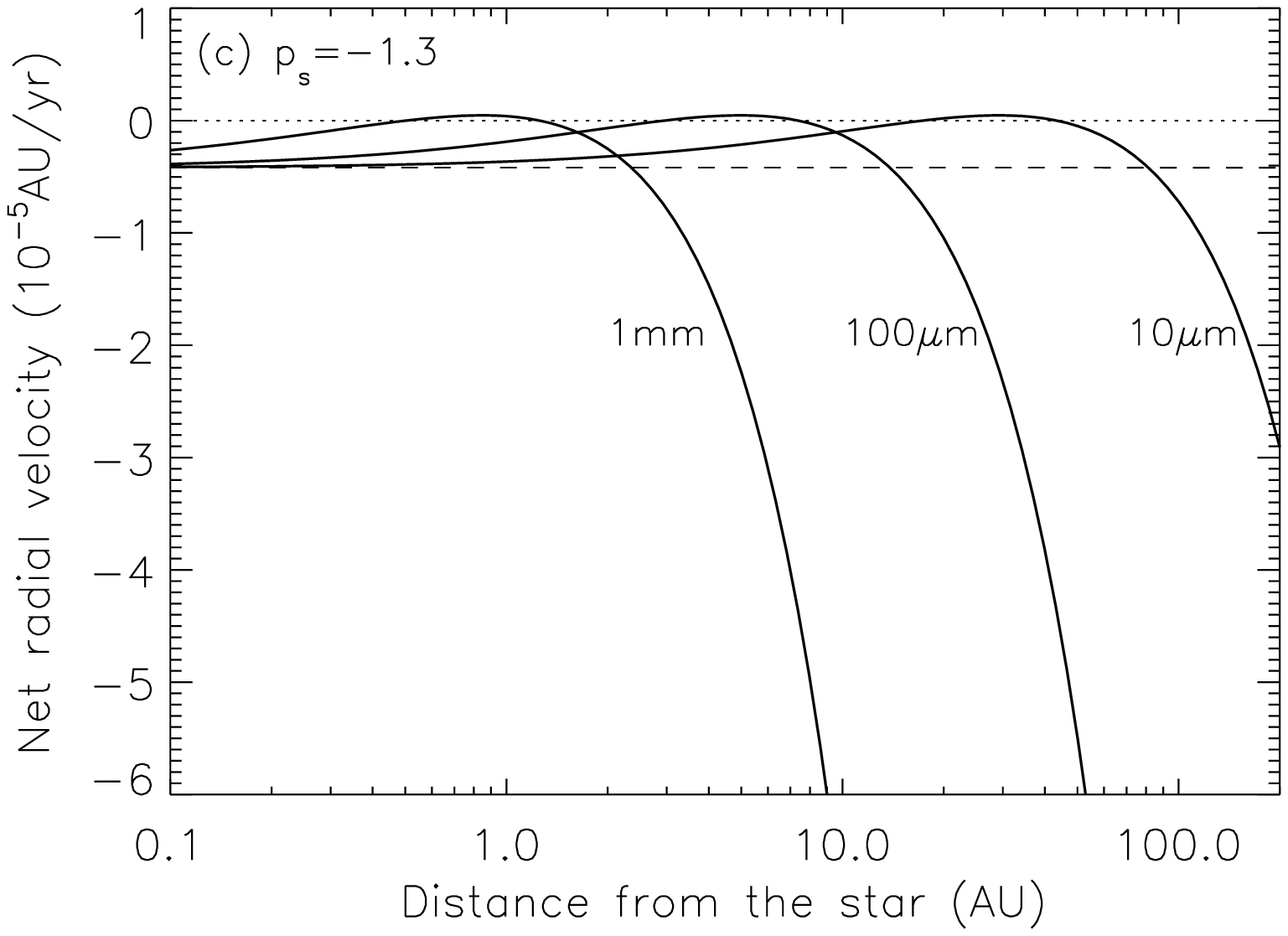}
\plotone{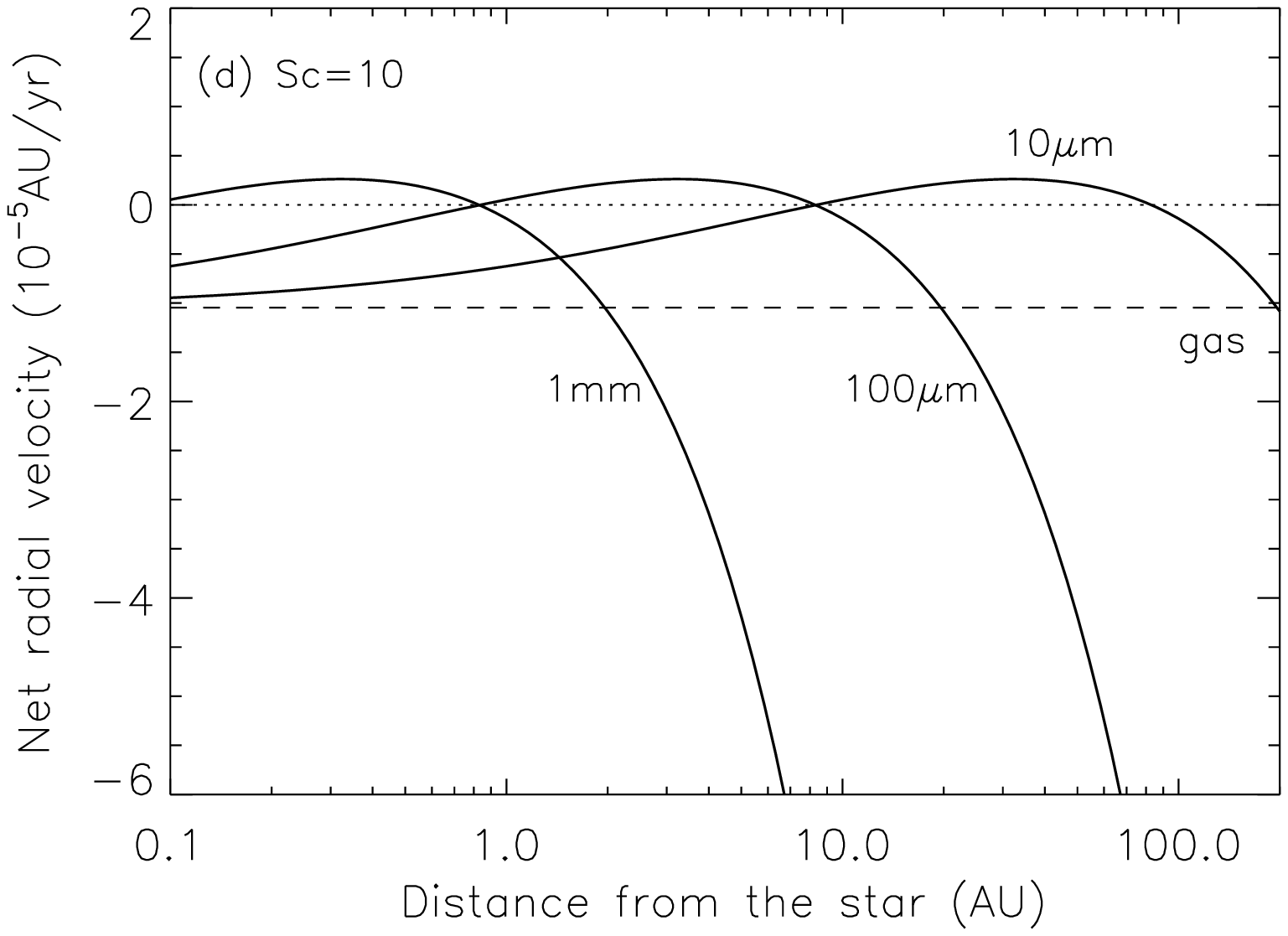}
\end{figure}
\clearpage

\begin{figure}
\epsscale{1.0}
\plotone{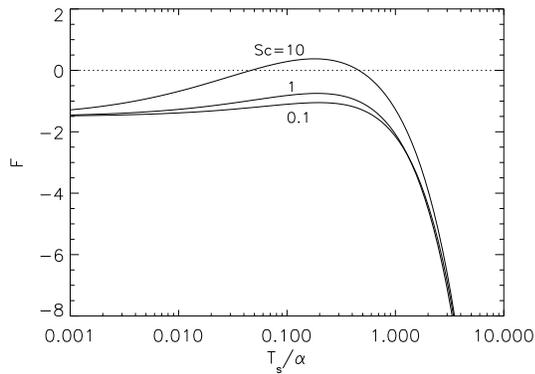}
\caption{
Function $F$ for various values of the Schmidt number, ${\rm Sc}=10$,
$1$, and $0.1$ from the upper line. The dotted line shows the zero for
reference.
\label{fig:f}
}
\end{figure}

\begin{figure}
\epsscale{1.0}
\plotone{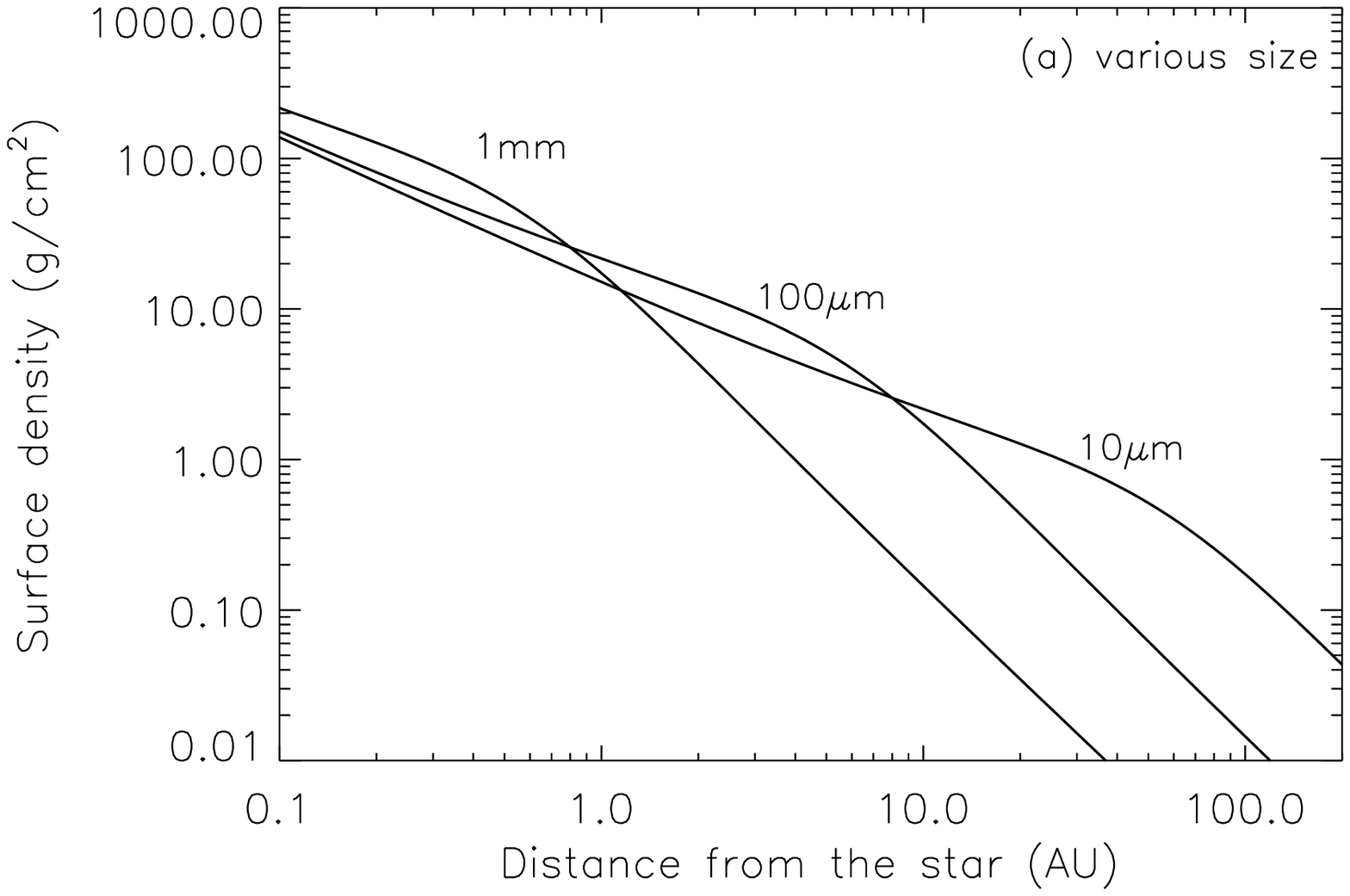}
\plotone{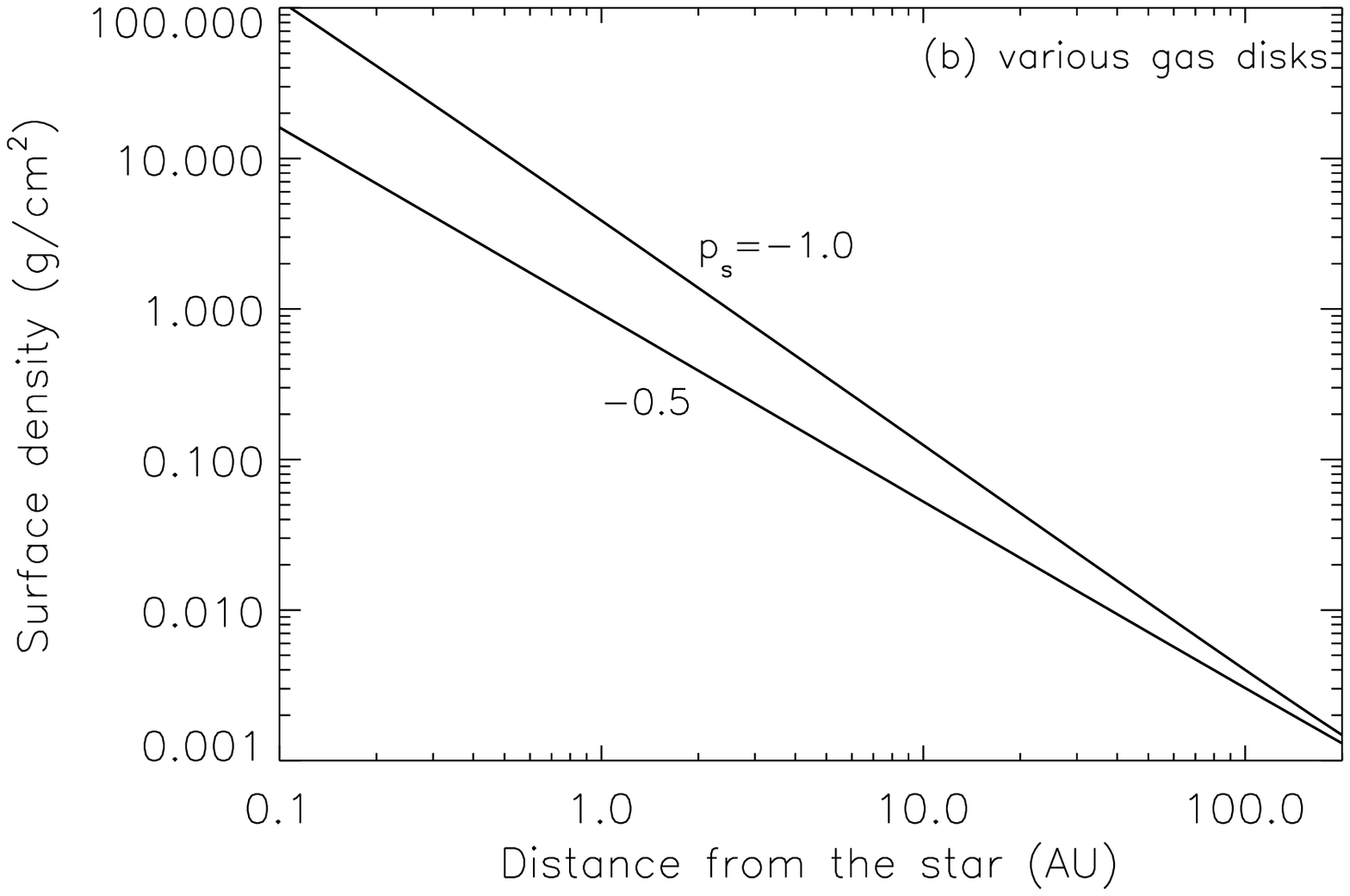}
\caption{
Surface density of dust disks, $\Sigma_d$.
($a$) The disk is composed of single sized particles of $s=1$ mm,
$100 \ \micron$, and $10 \ \micron$ from the left line. The mass flux
$\dot{M_d}$ is assumed to be $10^{-10} M_{\sun} \ {\rm yr}^{-1}$. (The
mass flux of the gas is $10^{-8} M_{\sun} \ {\rm yr}^{-1}$.)
The gas disk is the standard model ($p_s = -1.0$).
($b$) The disk is composed of particles with a power law size
distribution. The indices of the surface density profiles of the gas
disks are $p_s = -1.0$ and $-0.5$ from the upper line.
The particle number flux is assumed to be proportional to
$s^{-3.5}$. The minimum and maximum sizes of particles are $0.1 \
\micron$ and $10$ cm, respectively. The total mass flux is
$\dot{M}_{d,{\rm all}}=10^{-10} M_{\sun} \ {\rm yr}^{-1}$.
\label{fig:surden}
}
\end{figure}

\end{document}